# DNA self-assembly of single molecules with deterministic position and orientation

*Aleksandra K. Adamczyk[1], Teun A.P.M. Huijben[2], Miguel Sison[3], Andrea di Luca[4], Germán Chiarelli[1], Stefano Vanni[4], Sophie Brasselet[3], Kim Mortensen[2], Fernando D. Stefani[5,6]\*, Mauricio Pilo-Pais[1]\*, and Guillermo P. Acuna[1]\**

[1] Department of Physics, University of Fribourg, Chemin du Musée 3, Fribourg CH-1700, Switzerland.

[2] Department of Health Technology, Technical University of Denmark, Anker Engelunds Vej 101, 2800 Kongens Lyngby, Denmark.

[3] Aix Marseille Univ, CNRS, Centrale Marseille, Institut Fresnel, F-13013 Marseille, France

[4] Department of Biology, University of Fribourg, Chemin du Musée 10, Fribourg CH-1700, Switzerland.

[5] Centro de Investigaciones en Bionanociencias (CIBION), Consejo Nacional de Investigaciones Científicas y Técnicas (CONICET), Godoy Cruz 2390, C1425FQD Ciudad Autónoma de Buenos Aires, Argentina.

[6] Departamento de Física, Facultad de Ciencias Exactas y Naturales, Universidad de Buenos Aires, Güiraldes 2620, C1428EHA Ciudad Autónoma de Buenos Aires, Argentina

**Corresponding Authors**

\* E-mail: fernando.stefani@df.uba.ar (F.D.S.), mauricio.pilopais@unifr.ch (M.P.P.), guillermo.acuna@unifr.ch (G.P.A.)

**Abstract**

An ideal nanofabrication method should allow the organization of nanoparticles and molecules with nanometric positional precision, stoichiometric control and well-defined orientation. The DNA origami technique has evolved into a highly versatile bottom-up nanofabrication methodology that fulfils almost all of these features. It enables the nanometric positioning of molecules and nanoparticles with stoichiometric control, and even the orientation of asymmetrical nanoparticles along predefined directions. However, orienting individual molecules has been a standing challenge, mainly due to unspecific electrostatic interactions. Here, we show how single molecules, namely Cy5 and Cy3 fluorophores, can be incorporated in a DNA origami with controlled orientation by doubly linking them to oligonucleotide strands that are hybridized while leaving enough unpaired bases to induce a stretching force. Particularly, we explore the effects of leaving 0, 2, 4, 6, and 8 unpaired bases and find extreme orientations for 0 and 8 unpaired bases, corresponding to the molecules being perpendicular and parallel to the DNA double helix, respectively. We foresee that these results will expand the application field of DNA origami towards the fabrication of nanodevices involving a wide range of orientation-dependent molecular interactions, such as energy transfer, intermolecular electron transport, catalysis, exciton delocalization, or the electromagnetic coupling of a molecule to specific resonant nano-antennas modes.

**Main Text**

Over the last decade, the DNA origami technique[1] has been consolidated as a highly versatile method for the parallel self-assembly of nanostructures, providing high throughput and homogeneity using rather inexpensive consumables and equipment[2]. In addition to a high flexibility of geometric designs, DNA origami also serves as a nanometric breadboard to accommodate nanoparticles, biomolecules, and small molecules with positional and stoichiometric control[3–6]. Naturally, such level of fine nanofabrication is finding application in many different areas, which include nanoelectronics[7,8], drug delivery[9], sensing with nanopores[10,11] and enhanced catalysis[12] among others. Nanophotonics has been a particularly fertile field of application for DNA origami as they were used to organize metallic nanoparticles and optically active molecules at the nanoscale, enabling nanodevices that present enhanced fluorescence[13–16], single molecule SERS[17–20], artificial light harvesting[21–23], plasmon assisted FRET[24,25], multichomophoric FRET[26–28], energy transfer from molecules to metal films or graphene[29,30], molecular emission directivity[31,32], and strong coupling at room temperature[33,34]. Single non-spherical nanoparticles surface-functionalized with oligonucleotides can be positioned with predefined orientations on DNA origami through preferential DNA hybridization along certain directions, as it has been shown with gold nanorods[35] and triangular nanoplates[36]. Incorporating small molecules, such as organic fluorophores, with controlled orientation into DNA origami is of utmost interest since numerous phenomena exhibit a strong orientation dependency. For example, energy transfer, intermolecular electron transport, exciton delocalization, or the electromagnetic coupling of a molecule to specific resonant modes of nano-antennas, become maximum for a particular molecular orientation and negligible for the wrong orientation[37–39]. However, controlling both the position and orientation of small molecules in DNA origami has been a considerable challenge. In turn, this lack of control has limited the efficiency and reproducibility of experiments and applications. For example, a variety of photonic wires and 3D molecular networks have been assembled in DNA origami to gather and transmit photonic energy[21–23,27], but their efficiency is never going to match the natural light harvesting complexes without control over the orientation of the composing molecules.

Several strategies to incorporate small molecules in a DNA origami have been explored, with most of the work carried out with organic fluorophores. Using fluorophores is of interest not only due to their application in fluorescence imaging and assays down to the single molecule level, but also because fluorescence provides orientation-dependent signals. Analogously to the case of nanoparticles, small molecules linked to an oligonucleotide can be incorporated into specific positions of DNA origami by hybridization, even dynamically as it is done in DNA-PAINT[40]. In this case, the molecules present a high mobility. Alternatively, small molecules can bind noncovalently to double-stranded DNA (dsDNA) helices. This approach offers orientational control because molecules bind differently to dsDNA depending on their chemical structure. While some molecules bind preferentially in between bases (intercalators), others present higher affinity to the minor or major groove (groove binders), or to the external surface of the dsDNA chain[41]. Gopinath et al.[42] reported an example of this approach labelling DNA origami structures with the intercalating dye TOTO-3, which forms an angle of 70°± 10° with the axis of the dsDNA helix. Despite this accomplishment, this level of orientational control[43] comes at the expense of losing stoichiometric and positioning control because it is not possible to predefine the positions nor the number of binding molecules.

Yet another strategy to incorporate small molecules into a DNA origami consists of attaching them covalently to specific constituent single-stranded DNA (ssDNA) staples[44]. This approach provides high positional and stoichiometric control[26], but the resulting molecular orientation was unpredictable until now, partly due to the complex interaction between small molecules and DNA, which may depend on the molecular identity, the type of linker and the surrounding environment[45,46]. Also, determining the orientation of a single molecule with respect to DNA origami structures was made possible only recently[45].

Here, we present a study of the orientation of single Cy3 and Cy5 cyanine molecules relative to a DNA origami structure for different incorporation strategies and demonstrate that their position and orientation may be predictably controlled by modifying their linkage to the host structure. When these fluorophores, doubly linked to ssDNA staples, are hybridized in a specific manner to the origami leaving different numbers of unpaired bases, the molecular orientation can be controlled from being parallel to being perpendicular to the dsDNA helix-bundle in the origami.

Figures 1a and 1b show a sketch of the DNA origami employed. It consists of a two-layer 12-helix rectangular structure with dimensions of approximately 180 nm × 20 nm × 5 nm (length × width × height), with a protrusion at the center (further details about the DNA origami design are provided in Supplementary Note 1 and Figure S1). These structures were immobilized onto glass coverslips previously functionalized with neutravidin by means of 6 biotinylated ssDNA staples that define the underside (Figure 1a). On the upper side, 28 ssDNA staples were extended with a 20-nucleotide sequence to form three sites for DNA PAINT measurements (12 and 6 at the opposite ends, 10 at the center), with an average separation of 70 nm. Figure 1b schematically shows the strategy to incorporate single fluorophores with controlled orientations. A Cy5 or Cy3 fluorophore is doubly linked to a ssDNA staple (see Figure S2) with sequences designed to leave 0, 1, 2, 3, or 4 unpaired bases in the corresponding DNA scaffold where it hybridizes. In this way, the total number of unpaired bases at the fluorophore´s position could be varied between 0, 2, 4, 6, or 8 with the aim of studying the space required to stretch the molecule and restrain its mobility along the helix-bundle direction. The example in Figure 1b shows the case with a total of 4 unpaired bases. To ensure stability, the ssDNA staple was designed to have 12 base-pairs at each side of the fluorophore within the same helix bundle. In order to test the reproducibility of this approach, Cy5 and Cy3 molecules were incorporated at two different positions in the origami (Pos1 and Pos2 in Figure 1a) involving an equivalent geometry (see Figure 1b) but different DNA sequence.

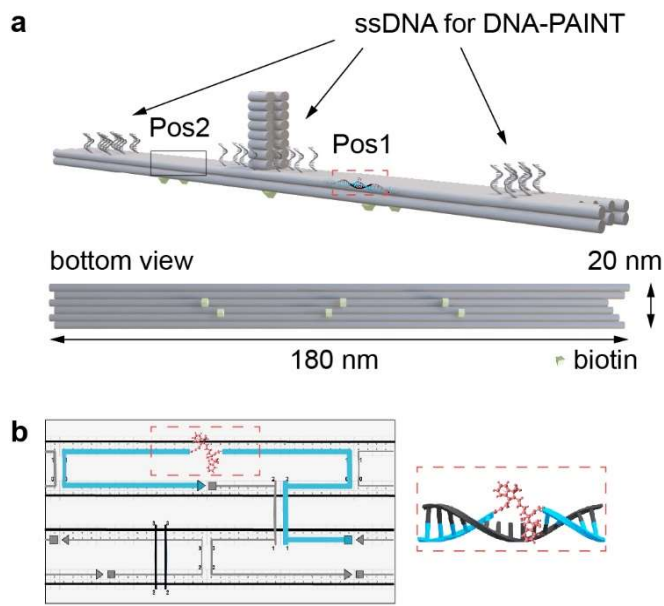

**Figure 1**: Sketch of the DNA origami host structure. (a) Perspective showing the three sites with ssDNA extensions for DNA PAINT and the two positions (Pos1 and Pos2) where the doubly linked fluorophores were incorporated (upper panel). Bottom view including the biotin modifications used for immobilization on the substrate (lower panel). (b) strategy to manipulate the orientation of molecules by leaving scaffold bases unpaired; zoom of the cADNano design (left, ssDNA staple modified with the doubly linked dye in light blue, adjacent ssDNA staples in gray and scaffold in black) showing four double-helices and zoom of the dye with the unpaired scaffold bases (right, in this example 4 in total).

The in-plane orientation of the fluorophore´s absorption transition dipole relative to the DNA origami ($\rho$, Figure 2a) was determined as previously reported[45]. Briefly, the orientation of each DNA origami with respect to the microscope coordinates ($\rho_{DNA}$) was determined through DNA-PAINT super-resolved imaging (Figure 2b). The orientation of the absorption transition dipole of the single fluorophores relative to the microscope coordinates ($\rho_F$) was obtained from polarization dependent excitation measurements (Figure 2c). From these two measurements, the orientation of the absorption transition dipole relative to the origami structure is obtained as $\rho = \rho_F - \rho_{DNA}$ (further details in Supplementary Note 2). Taking into account the dipolar symmetry and that the lowest electronic transition in carbocyanine dyes is polarized along the principal axis of the molecule[47–49], $\rho$ reports the in-plane projection of the molecular orientation as depicted in Figure 2a. Due to the symmetry of the DNA PAINT sites, we constrain our results in the range $\rho \in [0,90°]$, which is the relevant angle to characterize most orientation-dependent molecular interactions.

Measuring $\rho$ for a population of single molecules readily answers the question of whether the fluorophores present a preferential orientation within the DNA origami structure. In addition, the polarization-dependent excitation measurement also delivers the modulation ($M$) of the single molecule emission intensity as the excitation polarization is rotated (Figure 2c). The modulation $M$ provides information about the wobbling of the orientation (aperture δ, Figure 2a), taking a value from 1 for a molecule with a perfectly fixed in-plane orientation (δ ~ 0°), to 0 for a freely rotating molecule in a time scale faster than the polarization rotation (δ ~ 180°). The out-of-plane angle (η, Figure 2a) of a wobbling molecule further contributes to a decrease of $M$.

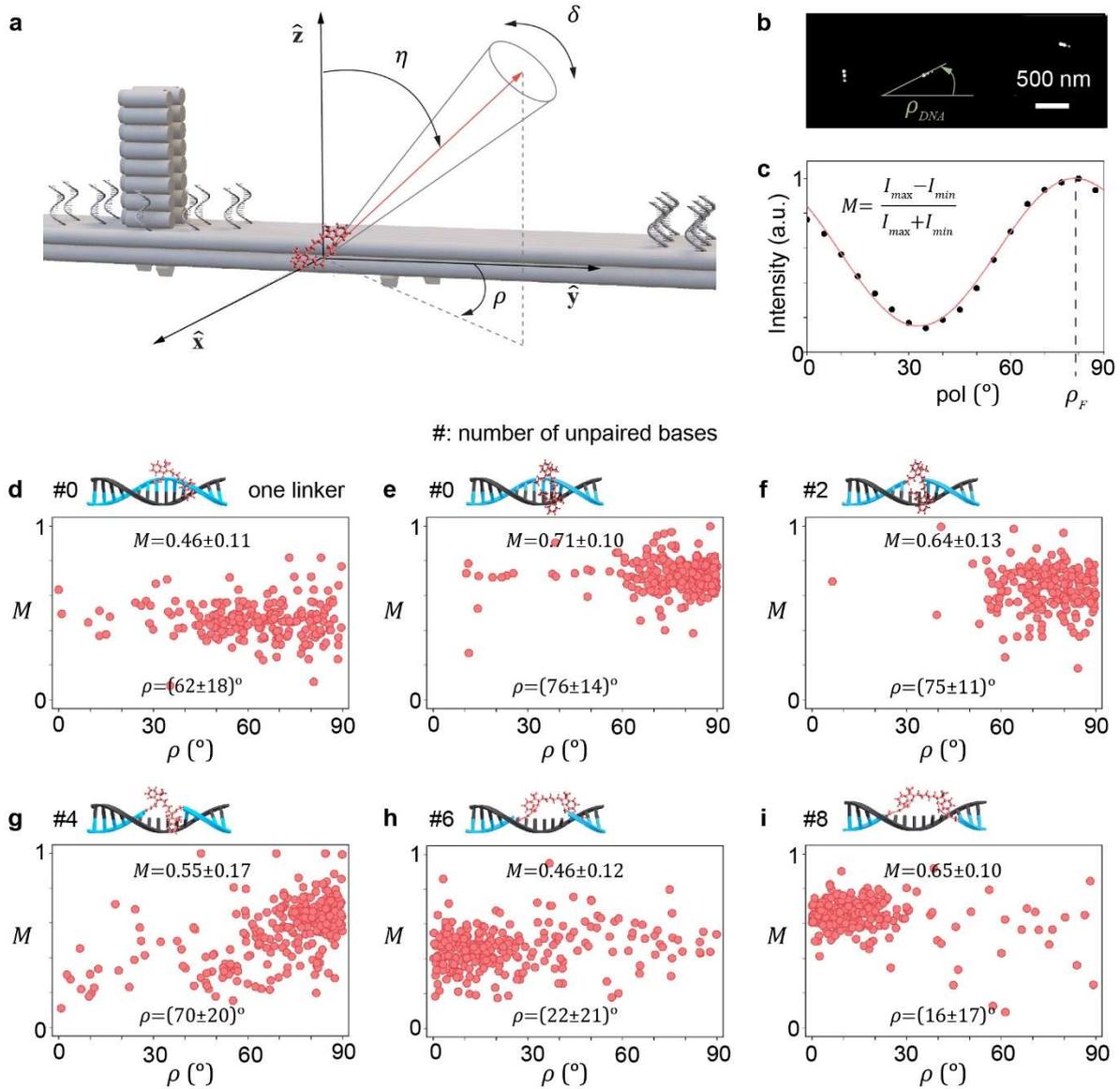

**Figure 2**. Molecular angular coordinates, in-plane angle $\rho$ and modulation $M$ for different binding schemes at Pos1. (a) Spherical coordinates system employed to describe the orientation and wobbling of the single fluorophores in the DNA origami. (b) Exemplary super-resolution image of DNA origamis obtained through DNA-PAINT measurements. The three super-resolved DNA-PAINT spots reveal the orientation of each DNA origami with respect to the microscope coordinates ($\rho_{DNA}$). (c) Single molecule fluorescence signal vs the incident polarization angle. The angle of maximum emission corresponds to the in-plane molecular orientation with respect to the microscope coordinates ($\rho_F$). (d-i) Scatter plots of $M$ vs. $\rho$ for Cy5 incorporated in Pos 1 through (d) one linker and 0 unpaired bases, (e-i) through a double link leaving 0 (e), 2 (f), 4 (g), 6 (h) and 8 (i) bases of the scaffold unpaired. The number of DNA origami structures studied in (d-i) was $n = 272, 229, 215, 263, 283, 247$ respectively.

Figure 2d shows a scatter plot of $M$ vs. $\rho$ for Cy5 molecules incorporated at Pos1 through an internal modification of the ssDNA staple using a single linker (see Figure S2). Cy5 molecules attached in this way present a rather broad distribution of $\rho$ ranging from 45° to 90° and an average modulation close to 0.5. The situation changes dramatically when the Cy5 molecule is doubly linked to the DNA origami (Figures 2e-i). For 0 bases unpaired, the molecules orient mostly perpendicularly to the host DNA double-helix with a high modulation (Figure 2c), whereas if 2 bases are left unpaired the in-plane orientation remains rather unchanged but lower values of $M$ are observed (Figure 2d). If 4 bases are left unpaired (Figure 2e), a much broader distribution of $M$ is observed and a considerable fraction of the fluorophores exhibit values of $\rho < 60°$. Finally, when 6 or 8 bases are left unpaired leading to a gap comparable to the extension of the dye with linkers (Figures 2f and 2g), the molecular orientation has transitioned from being perpendicular ($\rho \approx 90°$) to being parallel ($\rho \approx 0°$) to the dsDNA host. In particular, for 8 bases unpaired a high value of $M = 0.65 \pm 0.1$, comparable to that of 0 bases unpaired, is recovered.

In order to rationalize these findings, we claim that the DNA origami is capable of exerting a force due to the difference in persistence length between ssDNA and dsDNA, in analogy to previous reports[50]. Therefore, by leaving a number of bases unpaired, the fluorophore not only has enough space to fit within the host double-helix[51–53] but it is also stretched along the host double-helix by the DNA origami construct. To test this hypothesis, we performed further experiments. Figure 3a shows the results for Cy5 incorporated at the end of the ssDNA staple through one linker and with a redesign of the staples to also leave 8 scaffold bases unpaired. For this binding scheme, the Cy5 molecules present low values of $M \approx 0.4$ and a rather uniform distribution of $\rho$ in the 45° - 90° range, in agreement with previous reports[45]. These results confirm that the dye is not aligned along the host double helix due to sticking to the unpaired bases of the scaffold, and that the stretching from both ends of the doubly-linked fluorophore is required. The robustness of the approach was further evaluated by applying it with a different fluorophore (Cy3) and in an alternative position (Pos2, see Figure 1a) with the same binding scheme albeit another DNA sequence. The results obtained show consistently that both fluorophores align along the host double-helix when 8 bases are left unpaired (Figure 3b-d), independently of the position and sequence in the origami, whereas for 0 bases unpaired the fluorophores adopts a more perpendicular orientation (Figure S3).

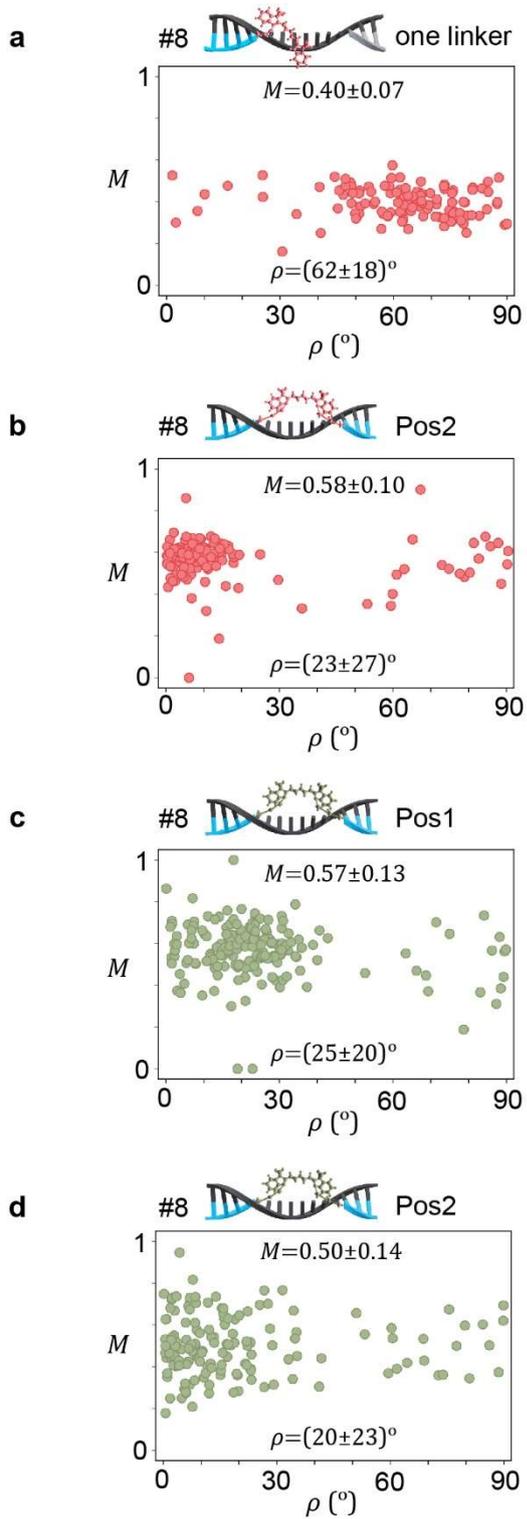

**Figure 3**. Scatter plots of $M$ vs. $\rho$ for Cy5 incorporated at the end of a ssDNA staple through one linker with the adjacent ssDNA staple shortened by 8 bases (a), and for doubly-linked Cy5 or Cy3 molecules with 8 scaffold bases left unpaired (b-c); (b) Cy5 at position 2, (c) Cy3 at position 1, and (d) Cy3 at positions 2. The number of DNA origami structures studied in (a-d) was $n = 115, 324, 179, 151$ respectively.

These results already demonstrate that the orientation of a single molecule can be controlled on a DNA origami by adjusting the linking configuration. While the differences in $M$ values in Figs 2 and 3 can be attributed to a mixture of effects due to wobbling and off-plane orientation contribution, these can be disentangled by measuring separately the molecular orientation and wobbling in three dimensions. To this end, two methods were employed, point-spread function (PSF) analysis (Supplementary Note 3)[54], and four-polarizations image splitting (Supplementary Note 4)[55]. These methods, which were performed on the same samples and using the same setup, provide in principle identical information but are complementary in terms of $(\rho, \eta, \delta)$ parameter space regions for which accuracy and precision are expected to be high (Supplementary Notes 3 and 4). Figure 4a shows a scatter plot of the out-of-plane angle $\eta$ vs the in-plane angle $\rho$ for Cy5 in position 1, with 0 and 8 bases unpaired, obtained with the PSF imaging method. These results confirm that the in-plane orientation of Cy5 shifts from perpendicular to parallel to the dsDNA when the number of unpaired bases increases from 0 to 8. In addition, they show that this transition is accompanied by a change of the out-of-plane orientation as the value of $\eta$ increases from $26° \pm 11°$ to $55° \pm 12°$, confirming that by leaving 8 scaffold bases unpaired the dye is stretched along the dsDNA helix orientation until exhibiting a close to in-plane orientation.

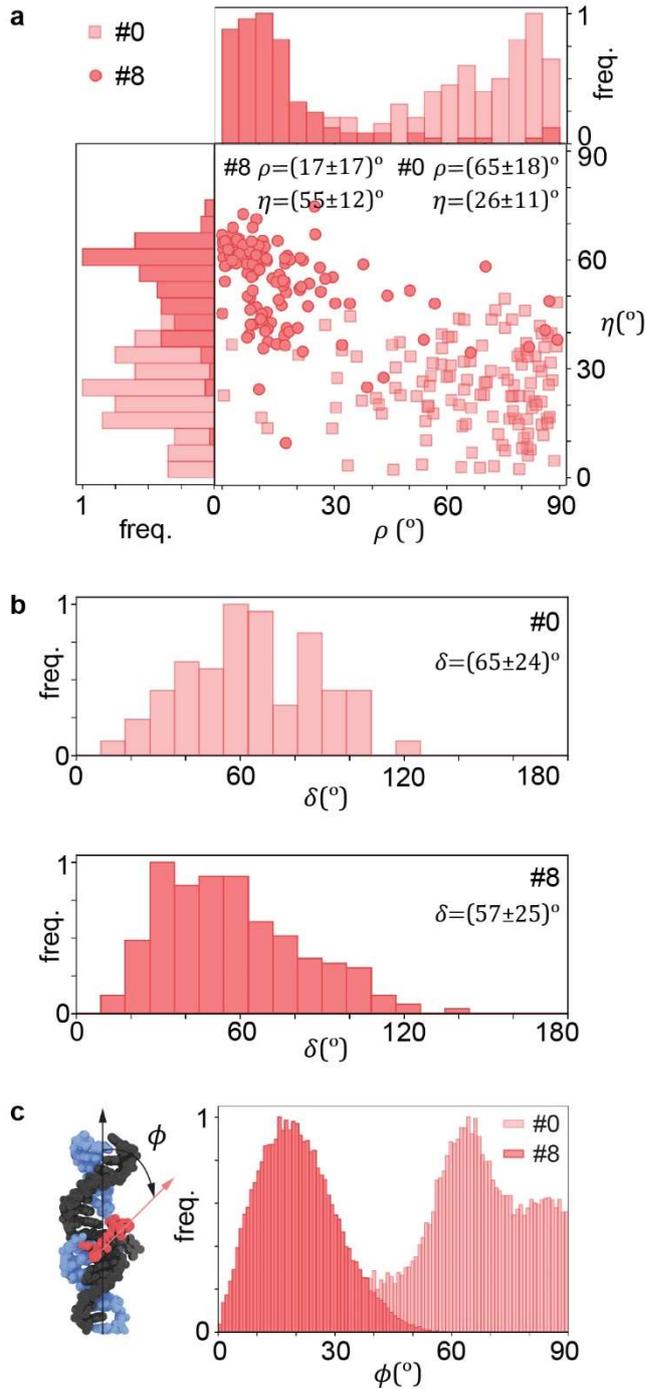

**Figure 4**. Three-dimensional orientation, wobbling and molecular dynamics (MD) simulations of Cy5 doubly linked in position 1 with 0 and 8 bases unpaired. (a) Scatter plot and distributions of $\eta$ vs. $\rho$. (b) Three-dimensional wobbling angle $\delta$. (c) Left: sketch of the model employed for the MD simulations (Cy5 in red, scaffold in black and ssDNA staple in blue). Right: histogram plot of the angle $\phi$ to the dsDNA helix ($\hat{y}$ axis) extracted from the MD simulations.

Figure 4b shows the distributions of $\delta$ for Cy5 doubly linked with 0 and 8 bases unpaired, obtained from four-polarizations image splitting measurements. In both linking configurations Cy5 present an average $\delta$ of around 60° with standard deviation of 25°. We note that this level of orientational control translates into variations of less than 10% for a dependency of $cos^2$ on molecular orientation.

Finally, we performed molecular dynamics simulations (Supplementary Note 5) of the extreme configurations for Cy5 and computed the angle with respect to the main axis of the double helix ($\phi$, Figure 4c). In agreement to the experimental observations, the simulations show that the most frequent position of the Cy5 molecules varies from being oriented close to the main axis of the double helix when 8 bases are unpaired bases to being nearly perpendicular for 0 unpaired bases (Figure 4c).

**Discussion**

Our results demonstrate that single small molecules such as organic fluorophores can be incorporated into predefined positions of a DNA origami with controlled orientation by adjusting their linkage conditions. The orientation of Cy5 and Cy3 molecules could be varied from being parallel to perpendicular with respect the double helix of dsDNA by linking the fluorophores from two sides with ssDNA and leaving different number of scaffold bases unpaired. While the molecular orientation with 0 unpaired bases may be influenced by interactions between the dye and the DNA, for 8 unpaired bases we could show that the orientation parallel to the dsDNA chain is obtained by pulling the fluorophore linkers. This, together with the fact that the orientational control was demonstrated at different positions of the origami, implies that other small molecules could also be oriented parallel to the dsDNA chains in these constructs.

In summary, our results consolidate DNA origami as a complete nanofabrication platform with full control at the molecular level. The capacity to organize small molecules individually with positional and orientational will push current and future applications of DNA origami to their maximum efficiency and reproducibility limit. For example, we foresee that future optical nanoantennas for single photon emitters, or synthetic multichromophoric light harvesting systems mimicking natural complexes based on DNA origami could be operated at maximum efficiency. Also, DNA origami with oriented molecules could provide orientation rulers to calibrate 3D polarized single molecule detection microscopy methods and assays. Another, less explored, field of application where placing molecules with controlled distance and orientation could be highly beneficial is catalysis. In analogy to enzymes, that speed up reactions by bringing reagents nearby with the correct relative orientation, future DNA origami with molecular orientational control could serve as synthetic catalytic platforms. Finally, we remark that the configurations we found to successful to control molecular orientation may not be the only ones, and further research should explore other possibilities.

**Author contributions**

F.D.S, M.P.P. and G.P.A. conceived the experiments. S.B., K.M., F.D.S, M.P.P. and G.P.A. designed the experiments. M.P.P. and A.K.A. designed and fabricated the DNA origami structures. G.C. modified the fluorescence microscope to perform the polarization resolved measurements. A.K.A, T.A.P.M.H. and M.S. performed the fluorescence measurements and analysed the results. S.V. and A.d.L. performed and analysed the numerical simulations. G.P.A. supervised the project. F.D.S. and G.P.A. wrote the manuscript with input from all the authors.

**Acknowledgements**

The authors thank Dr. Mathias Lakatos for fruitful discussion. This project has received funding from the European Union's Horizon 2020 research and innovation programme under the Marie Skłodowska-Curie grant agreement No 860914. SV acknowledges support by the Swiss National Science Foundation through the National Center of Competence in Research Bio-Inspired Materials. SV and ADL acknowledge support by the European Research Council under the European Union's Horizon 2020 research and innovation program (grant agreement no. 803952). This work was supported by grants from the Swiss National Supercomputing Centre under project ID s1030 and s1132. The research performed at I. Fresnel is funded by the "France 2030" French Government programs managed by the French National Research Agency (ANR-16-CONV-0001, ANR-21-ESRE-0002) and from the 3DPolariSR ANR grant (ANR-20-CE42-0003). F.D.S. acknowledges the support of the Max Planck Society and the Alexander von Humboldt Foundation. This work has been funded by Consejo Nacional de Investigaciones Científicas y Técnicas (CONICET) and Agencia Nacional de Promoción Científica y Tecnológica (ANPCYT), projects PICT-2017-0870, and PICT-2014-3729. G.P.A. acknowledges support from the Swiss National Science Foundation (200021_184687) and through the National Center of Competence in Research Bio-Inspired Materials (NCCR, 51NF40_182881).

*Nat. Commun.* **2022**, *13* (1), 301.